\documentclass[preprint,amsmath,amssymb,aps,showkeys,showpacs]{revtex4}
\usepackage[english]{babel}
\usepackage{graphicx}
\usepackage{graphics}
\usepackage{amsmath}
\usepackage{dcolumn}
\usepackage{amssymb}
\usepackage{bm}

\begin{document}
\title{Aharonov-Bohm effect for bound states in the cosmic string spacetime in the context of rainbow gravity}
\author{K. Bakke}
\email{kbakke@fisica.ufpb.br (Corresponding author)}
\affiliation{Departamento de F\'isica, Universidade Federal da Para\'iba, Caixa Postal 5008, 58051-900, Jo\~ao Pessoa, PB, Brazil.}

\author{H. Mota}
\email{hmota@fisica.ufpb.br}
\affiliation{Departamento de F\'isica, Universidade Federal da Para\'iba, Caixa Postal 5008, 58051-900, Jo\~ao Pessoa, PB, Brazil.}

\begin{abstract}

We explore a scenario of the general relativity determined by the framework of the rainbow gravity with the purpose of searching for analogues of the Aharonov-Bohm effect. By focusing on the confinement of the Dirac field and the scalar field to a hard-wall confining potential in a modified background of the cosmic string spacetime, we examine the effects of the rainbow gravity and the topology of the cosmic string spacetime. Then, we compare each spectrum of energy with the cases where the rainbow gravity is absent.

\end{abstract}


\maketitle

\section{Introduction}

The search for a consistent framework to understand and explain high-energy gravitational phenomena has been one of the main interest in theoretical physicists in the past few decades. Rainbow gravity, for instance, is a semi-classical approach which implies that the Lorentz symmetry is locally broken at energy scales comparable to the Planck scale $E_p$. It is in fact an extension of the so called Doubly Special Relativity \cite{Magueijo:2001cr, Magueijo:2002am, Magueijo:2002xx, AmelinoCamelia:2010pd, AmelinoCamelia:2000mn}. The modification of the metric, which depends on the ratio of the energy of a test particle to the Planck energy, is one of the main aspects of such framework and leads to high-order corrections to the energy-momentum dispersion relation. In fact, the suggestion that the relativistic dispersion relation should be modified has a motivation based on the observations of high-energy cosmic rays \cite{Magueijo:2001cr}, TeV photons from Gamma Ray Bursts \cite{Ellis, Abdo, Zhang} and IceCube data of neutrinos \cite{Amico}.

The programme of the rainbow gravity is based upon deformed Lorentz symmetries which makes possible to introduce, besides the invariance of the speed of light, an invariant length scale set as the Planck length, $\ell_p=\frac{1}{Ep}$ \cite{Magueijo:2001cr, Magueijo:2002am, Magueijo:2002xx, AmelinoCamelia:2010pd, AmelinoCamelia:2000mn}. Thus, the latter imposes an outset for the classical spacetime description to breakdown when length scales are smaller than $\ell_p$, leading to the spacetime to be thought of having a fuzzy microscopic structure. This feature is taken into consideration in proposals of a generalized uncertainty principle which makes possible to avoid singularities of general relativity \cite{mag1, mag2}. In the realm of rainbow gravity, the singularity associated with the beginning of the universe in the standard big-bang cosmology is also possible to be avoided \cite{Awad}. Also in the cosmological scenario, by using the rainbow gravity approach, the authors in Ref. \cite{nilsson} obtained lower bounds on the energy scale where the Lorentz symmetry is supposedly broken and showed that it can be as high as $10^{17}$\;Gev at 3$\sigma$ level, which is close to the Planck energy scale. Recently, the rainbow gravity has been considered in the context of quantum field theory \cite{Bezerra:2017zqq, Bakke:2018add, Bezerra:2019vrz,extra1,extra2}.

In this work, we focus on the semi-classical approach of the rainbow gravity, and thus, we analyse the Dirac and scalar fields in a modified background of the cosmic string spacetime. It is worth observing that, in the context of quantum field theory, cosmic strings have been considered in the past decades due to their gravitational, astrophysical and cosmological consequences \cite{Copeland:2011dx, Hindmarsh:2011qj}. It is a topological defect predicted to be formed in the early universe from the point of view of extensions of the Standard Model of particle physics and string theory. An idealized thin, straight and very long cosmic string, for instance, is characterized by spacetime with conical topology \cite{hindmarsh,VS}. This makes its gravitational field of great interest in quantum field theory in curved spacetime. Besides, the properties of a cosmic string in the framework of the rainbow gravity have been considered in Ref. \cite{Momeni:2017cvl}. Thereby, we search for a discrete spectrum of energy for both Dirac and scalar fields by confining them to a hard-wall confining potential in a modified background of the cosmic string spacetime yielded by the rainbow gravity. In addition, we analyse the relativistic analogues of the Aharonov-Bohm effect for bound states \cite{ABeffect}.

The structure of this paper is as follows: in section II, we introduce the scenario of the rainbow gravity, and thus, obtain the line element of the cosmic string spacetime in the context of rainbow gravity. in section III, we investigate the effects of rainbow gravity and the topology of the cosmic string on the Dirac field subject to a hard-wall confining potential; in section IV, we discuss the confinement of the scalar potential to a hard-wall confining potential under the effects of rainbow gravity  and the topology of the cosmic string; in section V, we present our conclusions.

\section{Modified cosmic string spacetime}

Let us introduce the cosmic string spacetime. It corresponds to a spacetime with a topological defect that is described by the line element \cite{Kibble:1976sj, Vilenkin:1984ib}:
\begin{eqnarray}
ds^{2}=-dt^{2}+dr^{2}+\eta^{2}\,r^{2}\,d\varphi^{2}+dz^{2}
\label{1.0}
\end{eqnarray}
where we have used the units $\hbar=1$ and $c=1$. The parameter $\eta$ is a constant related to the deficit angle. It is defined as $\eta=1-4\varpi$, where $\varpi$ is the linear mass density of the cosmic string. In the cosmic string spacetime, we have that $\eta<1$ \cite{Katanaev:1992kh,Furtado:1994nq}. Observe that $0\,<\,r\,<\,\infty$, $0\leq\varphi\leq2\pi$ and $-\infty\,<\,z\,<\,\infty$.

In the framework of rainbow gravity, the relativistic dispersion relation is modified due to the regime of high energy scale \cite{Magueijo:2001cr, Magueijo:2002am, Magueijo:2002xx, AmelinoCamelia:2010pd, AmelinoCamelia:2000mn, Bezerra:2017zqq, Bakke:2018add, Bezerra:2019vrz}. Thereby, for a relativistic particle with mass $m$ and momentum $p$, we can write
\begin{eqnarray}
E^{2}\,g_{0}^{2}\left(y\right)-p^{2}g_{1}^{2}\left(y\right)=m^{2}.
\label{1.1}
\end{eqnarray}
where $y=E/E_{p}$ is the ratio of the energy of the probe particle to the Planck energy $E_{P}$. Its meaning is that it regulates the level of the mutual relation between the spacetime background and the probe particles. Moreover, the function $g_{0}\left(y\right)$ and $g_{1}\left(y\right)$ are the rainbow functions. The behaviour of the rainbow function at low-energy regimes is 
\begin{eqnarray}
\lim_{y\rightarrow0}g_{k}\left(y\right)=1,
\label{1.2}
\end{eqnarray}
with $k=0,1$. Thereby, the framework of rainbow gravity permits us to write the line element of the cosmic string spacetime (\ref{1.0}) in the form \cite{Momeni:2017cvl, Bakke:2018add}:
\begin{eqnarray}
ds^{2}=-\frac{1}{g_{0}^{2}\left(y\right)}dt^{2}+\frac{1}{g_{1}^{2}\left(y\right)}\left[dr^{2}+\eta^{2}\,r^{2}\,d\varphi^{2}+dz^{2}\right].
\label{1.3}
\end{eqnarray}
Thus, the spacetime metric depends on the energy of the probe particle. This implies that particles with different frequencies (or wavelengths) will travel on different geodesics through the spacetime. The energy scale where this effect is expected to be observed is close to the Planck scale. For instance, by combining Hubble + Type Ia Supernovae + Baryon Acoustic Oscillations + Cosmic Microwave Background data, the authors in Ref. \cite{nilsson} obtained a lower bound on the energy of a massless particle which can be as high as $10^{17}$\;GeV. At this energy scale, deviations from the Lorentz invariance should be observed.

Henceforth, we shall work with a scenario determined by the rainbow functions \cite{Magueijo:2001cr, Magueijo:2002am, Magueijo:2002xx, AmelinoCamelia:2010pd, AmelinoCamelia:2000mn, Bezerra:2017zqq, Bakke:2018add, Bezerra:2019vrz}:
\begin{eqnarray}
g_{0}\left(y\right)=1;\,\,\,\,g_{1}\left(y\right)=\sqrt{1-\epsilon\,y^{2}}.
\label{1.4}
\end{eqnarray}
These rainbow functions have been considered previously to investigate their effects on the modification of the Friedmann-Robertson-Walker spacetime that describes the Standard big-bang cosmology \cite{Awad}. They have also been considered to study black hole evaporation due to Hawking radiation \cite{Ali}. It is worth mentioning that in most of quantum gravity theory proposals, the standard relativistic dispersion relation is modified. Rainbow gravity, although it is a semi-classical approach and not a fully-fledged theory of quantum gravity, follows the same direction in the attempt to investigate quantum gravity phenomena. Thereby, by using the rainbow gravity model described by the rainbow functions (\ref{1.4}) and in light of their interesting applications in several works as in \cite{Awad, Ali, Momeni:2017cvl, Magueijo:2001cr}, we have chosen to adopt the same rainbow function (\ref{1.4}) to investigate its effects on the energy levels of the Dirac and Klein-Gordon fields.

The rainbow functions (\ref{1.4}), when take into account in the line element (\ref{1.3}), provides
\begin{eqnarray}
ds^{2}=-dt^{2}+\frac{1}{\left(1-\epsilon\,y^{2}\right)}\left[dr^{2}+\eta^{2}\,r^{2}\,d\varphi^{2}+dz^{2}\right].
\label{1.4a}
\end{eqnarray}

Hence, Eq. (\ref{1.4a}) corresponds to the cosmic string spacetime in the context of the rainbow gravity described by the rainbow function (\ref{1.4}). Note that by taking $\epsilon\rightarrow0$, we recover the line element of the cosmic string spacetime (\ref{1.0}) \cite{Momeni:2017cvl}.

Since the line element (\ref{1.4}) describes a static spacetime and consequently there is no time evolution for the metric or for the rainbow functions, one could wonder whether it is possible to perform a re-scaling of the spacetime coordinates and end up with a line element free of rainbow functions, i.e., without rainbow gravity. This is in fact not possible because the conceptual notion of the measurement process is changed in the framework of rainbow gravity \cite{Kho}. Thereby, while such re-scaling is possible in practice, we would not obtain in the present case a genuine cosmic string spacetime free of rainbow gravity \cite{Kho}. Besides, the energy-dependent metric in the cosmological scenario seems to remove the big-bang singularity \cite{Awad, Kho} or solve the horizon problem \cite{Magueijo:2001cr}, hence, it indicates that the rainbow gravity has indeed real physical consequences.

It is worth mentioning that the Doubly (or Deformed) Special Relativity scenario, from which the rainbow gravity is extended, naturally arises from the asymptotic safety gravity \cite{xavier}. In the latter, the authors showed that the emergent Deformed Special Relativity has a key difference, i.e., it depends on the scattering process of particles considered. Moreover, the authors of Ref. \cite{MA} also discuss the emergence of an energy-dependent metric from a quantum cosmological model built in the context of quantum field theory. On the other hand, the author of Ref. \cite{sabine} showed that the requirement that an energy-dependent speed of light should be observer independent leads to a violation of locality. Photons from distant gamma-ray bursts suggested that, in order for this non-locality not be in contrast with particle interactions already measured, the energy-dependence of the speed of light cannot be of first order in the energy over the Planck mass. Bounds on possible modifications of the speed of light was then derived from this conclusion and ruled out the first order dependence.

Despite several issues of rainbow gravity still need to be addressed, such as, the understanding of how it fits with the standard model of particle physics, how the quantization works in its realm and the troubles with its non-locality feature \cite{sabine}, some progress seems to have been made and shows that rainbow gravity may be plausible as an emergent energy-dependent metric approach from other quantum gravity theory proposals \cite{xavier, MA}. Hence, we think it is worthwhile trying to investigate how spinor and boson fields have their energy levels affected by the energy-dependent metric from rainbow gravity.

\section{Dirac field in the cosmic string in the context of rainbow gravity }

In this section, we deal with the Dirac field in the cosmic string spacetime in the context of the rainbow gravity. By using the spinor theory in curved space \cite{birrell1984quantum,nakahara2003geometry}, we need to introduce a noncoordinate basis $\hat{\theta}^{a}=e^{a}_{\,\,\,\mu}\left(x\right)\,dx^{\mu}$. The components $e^{a}_{\,\,\,\mu}\left(x\right)$ are known as tetrads and give rise to the local reference frame of the observers. They satisfy the relation:
\begin{eqnarray}
g_{\mu\nu}\left(x\right)=e^{a}_{\,\,\,\mu}\left(x\right)\,e^{b}_{\,\,\,\nu}\left(x\right)\,\eta_{ab},
\label{1.5}
\end{eqnarray}
where $\eta_{ab}=\mathrm{diag}\left(-\,+\,+\,+\right)$ is the Minkowski metric. Moreover, the inverse of the tetrads is given by $dx^{\mu}=e^{\mu}_{\,\,\,a}\left(x\right)\hat{\theta}^{a}$, where $e^{a}_{\,\,\,\mu}\left(x\right)e^{\mu}_{\,\,\,b}\left(x\right)=\delta^{a}_{b}$. Therefore, from the line element (\ref{1.4a}), we can write
\begin{eqnarray}
\hat{\theta}^{0}=dt;\,\hat{\theta}^{1}=\frac{1}{\sqrt{1-\epsilon\,y^{2}}}\,dr;\,\,\,\hat{\theta}^{2}=\frac{\eta\,r}{\sqrt{1-\epsilon\,y^{2}}}\,d\varphi;\,\,\,\hat{\theta}^{3}=\frac{1}{\sqrt{1-\epsilon\,y^{2}}}\,dz.
\label{1.6}
\end{eqnarray}

Next, we can solve the Maurer-Cartan structure equations in the absence of torsion \cite{nakahara2003geometry}: $d\hat{\theta}^{a}+\omega_{\mu\,\,\,\,b}^{\,\,\,a}\left(x\right)dx^{\mu}\wedge\,\hat{\theta}^{b}$. Thus, we have $\omega_{\varphi\,21}\left(x\right)=-\omega_{\varphi\,12}\left(x\right)=-\eta$. With the purpose of writing the Dirac equation in the background described by the line element (\ref{1.4a}), we need to obtain the spinorial connection. It is defined as $\Gamma_{\mu}\left(x\right)=\frac{i}{4}\omega_{\mu\,ab}\left(x\right)\,\Sigma^{ab}$, where $\Sigma^{ab}=\frac{i}{2}\left[\gamma^{a},\,\gamma^{b}\right]$. Note that the mathematical object $\gamma^{a}$ corresponds to the Dirac matrices in the Minkowski spacetime \cite{greiner1990relativistic}:
\begin{eqnarray}
\gamma^{0}=\hat{\beta}=\left(
\begin{array}{cc}
1 & 0 \\
0 & -1 \\
\end{array}\right);\,\,\,\,\,\,
\gamma^{i}=\hat{\beta}\,\hat{\alpha}^{i}=\left(
\begin{array}{cc}
 0 & \sigma^{i} \\
-\sigma^{i} & 0 \\
\end{array}\right);\,\,\,\,\,\,\Sigma^{i}=\left(
\begin{array}{cc}
\sigma^{i} & 0 \\
0 & \sigma^{i} \\	
\end{array}\right).
\label{1.7}
\end{eqnarray}
The matrix $\vec{\Sigma}$ is the spin vector. Besides, the matrices $\sigma^{i}$ are the Pauli matrices. They satisfy the relation $\frac{1}{2}\left(\sigma^{i}\,\sigma^{j}+\sigma^{j}\,\sigma^{i}\right)=\eta^{ij}$. Hence, the only one non-null component of the spinorial connection is $\Gamma_{\varphi}=-\frac{\eta}{2}\,\Sigma^{3}$ \cite{Bakke:2018add}.

Therefore, in the context of curved space, the covariant form of the Dirac equation is given by
\begin{eqnarray}
m\psi=i\gamma^{\mu}\partial_{\mu}\psi+i\gamma^{\mu}\Gamma_{\mu}\left(x\right)\psi,
\label{1.8}
\end{eqnarray}
where the relation between the $\gamma^{\mu}$ matrices with the $\gamma^{a}$ matrices is given by $\gamma^{\mu}=e^{\mu}_{\,\,\,a}\left(x\right)\gamma^{a}$ \cite{birrell1984quantum}. In this way, by using (\ref{1.6}), the Dirac equation becomes
\begin{eqnarray}
i\frac{\partial\psi}{\partial t}&=&m\hat{\beta}\psi-i\sqrt{1-\epsilon\,y^{2}}\,\gamma^{0}\gamma^{1}\left(\frac{\partial}{\partial r}+\frac{1}{2r}\right)\psi-i\frac{\sqrt{1-\epsilon\,y^{2}}}{\eta\,r}\,\gamma^{0}\gamma^{2}\,\frac{\partial\psi}{\partial\varphi}\nonumber\\
&-&i\sqrt{1-\epsilon\,y^{2}}\,\gamma^{0}\gamma^{3}\frac{\partial\psi}{\partial z}.
\label{1.9}
\end{eqnarray}
Then, the solution to the Dirac equation (\ref{1.9}) can be given in the form:
\begin{eqnarray}
\psi=e^{-iEt}\left(
\begin{array}{c}
\chi_{1}\\
\chi_{2}\\
\end{array}\right),
\label{1.10}
\end{eqnarray}
where $\chi_{1}$ and $\chi_{2}$ are spinors of two-components \cite{Bakke:2018add,Bakke:2013wla, Bakke:2013jga}. By substituting (\ref{1.10}) into the Dirac equation (\ref{1.9}), we obtain two coupled equations for $\chi_{1}$ and $\chi_{2}$. The first one is 
\begin{eqnarray}
\left[\frac{E-m}{\sqrt{1-\epsilon\,y^{2}}}\right]\chi_{1}=-i\sigma^{1}\left[\frac{\partial}{\partial r}+\frac{1}{2r}\right]\chi_{2}-\frac{i\sigma^{2}}{\eta\,r}\frac{\partial\chi_{2}}{\partial\varphi}-i\sigma^{3}\frac{\partial\chi_{2}}{\partial z}.
\label{1.11}
\end{eqnarray}
The second coupled equation is
\begin{eqnarray}
\left[\frac{E+m}{\sqrt{1-\epsilon\,y^{2}}}\right]\chi_{2}=-i\sigma^{1}\left[\frac{\partial}{\partial r}+\frac{1}{2r}\right]\chi_{1}-\frac{i\sigma^{2}}{\eta\,r}\,\frac{\partial\chi_{1}}{\partial\varphi}-i\sigma^{3}\frac{\partial\chi_{1}}{\partial z}.
\label{1.12}
\end{eqnarray}
By eliminating $\chi_{2}$ in Eq. (\ref{1.12}), and thus, by substituting it in Eq. (\ref{1.11}), we obtain the second-order differential equation for $\chi_{1}$:
\begin{eqnarray}
\left[\frac{E^{2}-m^{2}}{\left(1-\epsilon\,y^{2}\right)}\right]\chi_{1}=-\frac{\partial^{2}\chi_{1}}{\partial r^{2}}-\frac{1}{r}\frac{\partial\chi_{1}}{\partial r}-\frac{1}{\eta^{2}\,r^{2}}\,\frac{\partial^{2}\chi_{1}}{\partial\varphi^{2}}+\frac{i\,\sigma^{3}}{\eta\,r^{2}}\,\frac{\partial\chi_{1}}{\partial\varphi}+\frac{1}{4r^{2}}\chi_{1}-\frac{\partial^{2}\chi_{1}}{\partial z^{2}}.
\label{1.13}
\end{eqnarray}

We can observe that $\chi_{1}$ is an eigenfunction of $\sigma^{3}$. Therefore, $\sigma^{3}\chi_{1}=\pm\chi_{1}=s\chi_{1}$, where $\chi_{1}=\left(\chi_{1}^{+}\,\chi_{1}^{-}\right)^{\mathrm{T}}$. Besides, due to the cylindrical symmetry, we can write
\begin{eqnarray}
\chi_{1}=e^{i\left(l+1/2\right)\varphi+ikz}\,\left(
\begin{array}{c}
u_{+}\left(r\right)\\
u_{-}\left(r\right)\\
\end{array}\right),
\label{1.14}
\end{eqnarray}
where $l=0,\pm1,\pm2,\ldots$ and $k$ is a constant. From now on, let us simplify our discussion by taking $k=0$. Thus, by substituting (\ref{1.14}) into Eq. (\ref{1.13}), we have that the second-order differential equations for $u_{+}\left(r\right)$ and $u_{-}\left(r\right)$ are given by
\begin{eqnarray}
u_{s}''+\frac{1}{r}\,u_{s}'-\frac{\nu^{2}}{\eta^{2}\,r^{2}}\,u_{s}+\lambda^{2}\,u_{s}=0.
\label{1.15}
\end{eqnarray}
It is known in the literature as the Bessel differential equation \cite{Arfken,Abramowitz}. Besides, we have defined the parameters $\nu$ and $\lambda$ in Eq. (\ref{1.15}) as
\begin{eqnarray}
\nu&=&l+\frac{1}{2}\left(1-s\right)+\frac{s}{2}\left(1-\eta\right);\nonumber\\
[-2mm]\label{1.16}\\[-2mm]
\lambda^{2}&=&\left[\frac{E^{2}-m^{2}}{\left(1-\epsilon\,y^{2}\right)}\right].\nonumber
\end{eqnarray}
Hence, the solution to Eq. (\ref{1.15}) is given in the form:
\begin{eqnarray}
u_{s}\left(r\right)=A\,J_{\frac{\left|\nu\right|}{\eta}}\left(\lambda\,r\right)+B\,N_{\frac{\left|\nu\right|}{\eta}}\left(\lambda\,r\right),
\label{1.17}
\end{eqnarray} 
where $A$ and $B$ are constants. The functions $J_{\frac{\left|\nu\right|}{\eta}}\left(\lambda\,r\right)$ and $N_{\frac{\left|\nu\right|}{\eta}}\left(\lambda\,r\right)$ are the Bessel functions of first and second kinds \cite{Arfken,Abramowitz}, respectively. We search for a regular solution at the origin. Therefore, it is obtained by taking $B=0$ in Eq. (\ref{1.17}), because the function $N_{\frac{\left|\nu\right|}{\eta}}\left(\lambda\,r\right)$ diverges at the origin. Then, the solution to Eq. (\ref{1.15}) becomes
\begin{eqnarray}
u_{s}\left(r\right)=A\,J_{\frac{\left|\nu\right|}{\eta}}\left(\lambda\,r\right).
\label{1.18}
\end{eqnarray}

\begin{figure}[!htb]
\begin{center}
\includegraphics[width=0.4\textwidth]{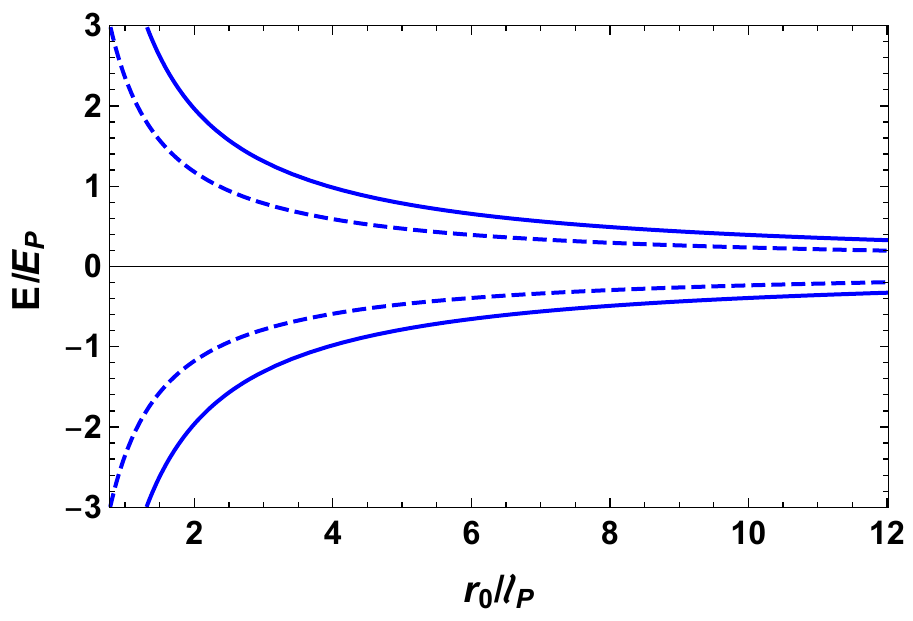}
%
\includegraphics[width=0.415\textwidth]{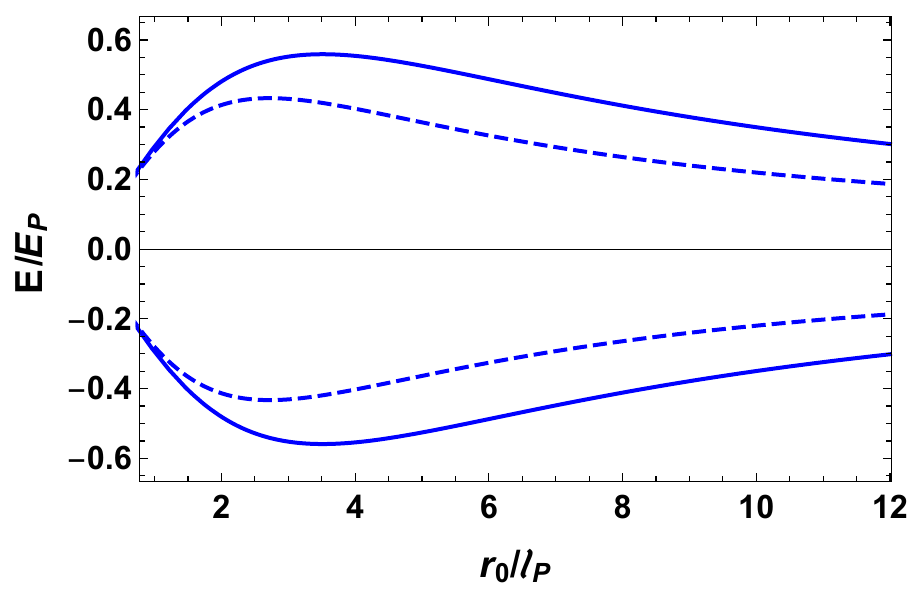}
\caption{\small{Plot of the energy levels \eqref{1.21}, in units of Planck's energy, in terms of the ratio of the radius $r_0$ to the Planck's length $\ell_p$, for $n=l=0$. The plot on the left is for $\epsilon=0$ while the one on the right is for $\epsilon=1$. For both plots we consider $\frac{m}{E_P}=0.01$, $\eta=1$. The dashed and full blue lines are for $s=+1$ and  $s=-1$, respectively. Note that the range of the dimensionless parameter $\frac{r_0}{\ell_p}$ goes from $1$ to $\infty$.}}
\label{f1}
\end{center}
\end{figure}

Let us consider the Dirac particle is confined to a hard-wall confining potential. For a fixed radius $r_{0}$, then, we have the boundary condition:
\begin{eqnarray}
u_{s}\left(r_{0}\right)=0.
\label{1.19}
\end{eqnarray}
Let us assume that $\lambda\,r_{0}\gg1$. Therefore, for $\lambda\,r_{0}\gg1$, when $\frac{\left|\nu\right|}{\eta}$ is fixed, the Bessel function can be written as \cite{Arfken,Abramowitz, Bezerra:1997mn}:
\begin{eqnarray}
J_{\left|\zeta\right|}\left(x_{0}\right)\rightarrow\sqrt{\frac{2}{\pi\,x_{0}}}\,\cos\left(x_{0}-\frac{\left|\zeta\right|\,\pi}{2}-\frac{\pi}{4}\right).
\label{1.20}
\end{eqnarray}

Thereby, by using the relation (\ref{1.20}) in Eq. (\ref{1.18}) (where we must replace $\left|\zeta\right|$ with $\frac{\left|\nu\right|}{\eta}$, and also $x_{0}$ with $\lambda\,r_{0}$), we obtain from the boundary condition (\ref{1.19}):
\begin{eqnarray}
E_{n,\,l,\,s}=\pm\sqrt{\frac{m^{2}+\frac{\pi^{2}}{r_{0}^{2}}\left[n+\frac{\left|\nu\right|}{2\eta}+\frac{3}{4}\right]^{2}}{\left[1+\frac{\epsilon\,\pi^{2}}{E_{p}^{2}\,r_{0}^{2}}\left(n+\frac{\left|\nu\right|}{2\eta}+\frac{3}{4}\right)^{2}\right]}  }.
\label{1.21}
\end{eqnarray}

Therefore, Eq. (\ref{1.21}) corresponds to the relativistic energy levels for the Dirac field subject to a hard-wall confining potential in the modified background of the cosmic string spacetime. By comparing with Ref. \cite{Bezerra:1997mn}, we have that the relativistic spectrum of energy of the Dirac field subject to a hard-wall confining potential changes due to the effects of rainbow gravity. The contribution to the relativistic energy level that stems from the effects of the rainbow gravity is given by the term $\left[1+\frac{\epsilon\,\pi^{2}}{E_{p}^{2}\,r_{0}^{2}}\left(n+\frac{\left|\nu\right|}{2\eta}+\frac{3}{4}\right)^{2}\right]^{-1}$ that appears inside the square root in Eq. (\ref{1.21}). On the other hand, by taking $\epsilon\rightarrow0$ in Eq. (\ref{1.21}), we recover the relativistic energy levels in the absence of rainbow gravity \cite{Bezerra:1997mn}.

Moreover, observe that the relativistic energy levels (\ref{1.21}) depend on the parameter $\eta$ that characterizes the topological defect spacetime. Besides, the effects of the topology of the spacetime gives rise to an effective angular momentum $l_{\mathrm{eff}}=\frac{\nu}{\eta}=\frac{1}{\eta}\left[l+\frac{1}{2}\left(1-s\right)+\frac{s}{2}\left(1-\eta\right)\right]$. The meaning of this effective angular momentum can be understood from the work of Peshkin and Tonomura \cite{ABeffect}. They showed when a point charge is confined to move in a circular ring of radius $R$ and there is a long solenoid of radius $a\,<\,R$ concentric to the ring, thus, the angular momentum quantum number is modified by $l'=l-e\Phi/2\pi$ (where $\Phi$ is the magnetic flux through the solenoid and $e$ is the electric charge). Furthermore, after obtained the eigenvalues of energy, they showed that the spectrum of energy is determined by $l'=l-e\Phi/2\pi$ even though no interaction between the point charge and the magnetic field inside the solenoid exists. This influence of the magnetic flux on the eigenvalues of energy is called as the Aharonov-Bohm effect for bound states. By returning to the effective angular momentum $l_{\mathrm{eff}}=\frac{\nu}{\eta}=\frac{1}{\eta}\left[l+\frac{1}{2}\left(1-s\right)+\frac{s}{2}\left(1-\eta\right)\right]$, hence, there is a shift in the angular momentum quantum number analogous to $l'=l-e\Phi/2\pi$ obtained by Peshkin and Tonomura \cite{ABeffect}. Therefore, even though no interaction between the Dirac field and topological defect exists, there is the presence of the effective angular momentum $l_{\mathrm{eff}}$ in the relativistic energy levels. This corresponds to an analogous effect to the Aharonov-Bohm effect for bound states \cite{ABeffect, Bezerra:1997mn}.

\begin{figure}[!htb]
\begin{center}
\includegraphics[width=0.45\textwidth]{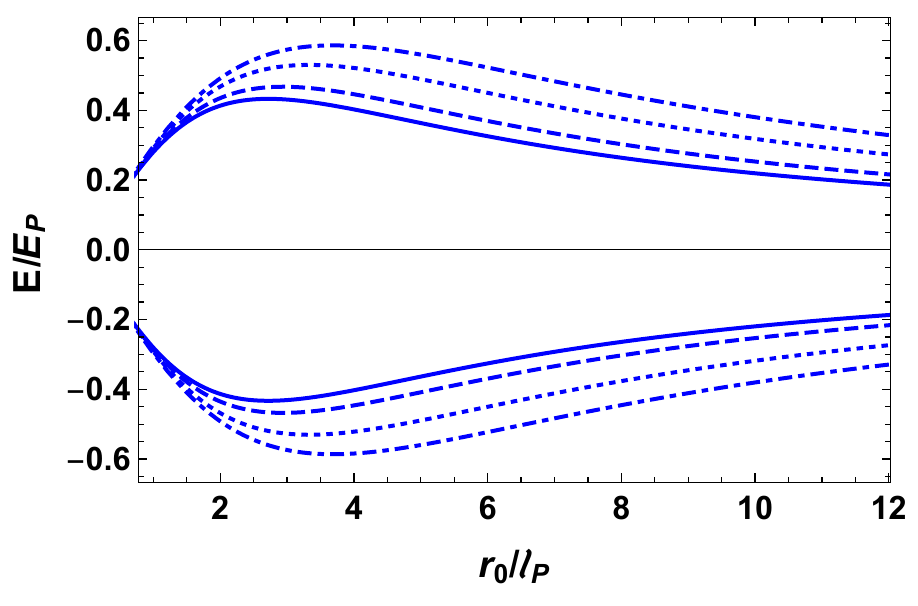}
%
\caption{\small{Plot of the energy levels \eqref{1.21}, in units of Planck's energy, in terms of the ratio of the radius $r_0$ to the Planck's length $\ell_p$, for $n=l=0$. We consider $\epsilon=1$, $\frac{m}{E_P}=0.01$ and $s=+1$. The full, dashed, dotted and dot-dashed lines are, respectively, for $\eta=1$, $\frac{1}{\eta}=1.5$, $\frac{1}{\eta}=2.5$ and $\frac{1}{\eta}=3.5$. Note that the range of the dimensionless parameter $\frac{r_0}{\ell_p}$ goes from 1 to $\infty$.}}
\label{f2}
\end{center}
\end{figure}

On the other hand, by taking $\eta\rightarrow1$, we have that the relativistic energy levels (\ref{1.21}) becomes
\begin{eqnarray}
E_{n,\,l,\,s}=\pm\sqrt{\frac{m^{2}+\frac{\pi^{2}}{r_{0}^{2}}\left[n+\frac{\left|\bar{\nu}\right|}{2}+\frac{3}{4}\right]^{2}}{\left[1+\frac{\epsilon\,\pi^{2}}{E_{p}^{2}\,r_{0}^{2}}\left(n+\frac{\left|\bar{\nu}\right|}{2}+\frac{3}{4}\right)^{2}\right]}},
\label{1.22}
\end{eqnarray}
where $\bar{\nu}=l+\frac{1}{2}\left(1-s\right)$. Therefore, Eq. (\ref{1.22}) corresponds to the relativistic energy levels for the Dirac field subject to a hard-wall confining potential in the modified background of the Minkowski spacetime. The effects of the rainbow gravity yield the contribution to the relativistic energy levels (\ref{1.22}) given by the term $\left[1+\frac{\epsilon\,\pi^{2}}{E_{p}^{2}\,r_{0}^{2}}\left(n+\frac{\left|\bar{\nu}\right|}{2}+\frac{3}{4}\right)^{2}\right]^{-1}$ that appears inside the square root. We should also observe that we can recover the relativistic energy levels in the absence of rainbow gravity \cite{Bezerra:1997mn} by taking $\epsilon\rightarrow0$ in Eq. (\ref{1.22}).

We have plotted, in Figs. \ref{f1} and \ref{f2}, the dimensionless energy levels (\ref{1.21}) in terms of the ratio $\frac{r_0}{\ell_p}$ by considering the ground state of the fermionic field, that is, $n=0$ and $l=0$. In Fig. \ref{f1}, on the left, we have plotted the dimensionless ground state energy in the absence of both rainbow gravity and cosmic string while, on the right, we consider the plot only due to the rainbow gravity modification, i.e., $\epsilon =1$. For these plots, we take $\frac{m}{E_P}=0.01$, $s=1$ (blue dashed lines) and $s=-1$ (full blue lines). It is clear that the introduction of the rainbow functions drastically modifies the energy for values of the radius $r_0$ close to the Planck length $\ell_p$ while for large values of $r_0$ the behaviour of the ground state energy tends to be the same as in the plot on the left.

As to the Fig. \ref{f2}, we have also plotted the dimensionless energy levels (\ref{1.21}) in terms of the ratio $\frac{r_0}{\ell_p}$ by considering the ground state of the fermionic field, that is, $n=0$ and $l=0$. However, in this case, we take $\frac{m}{E_P}=0.01$, $s=1$ and varies the cosmic string parameter $\eta$. The full, dashed, dotted and dot-dashed lines are, respectively, for $\eta=1$, $\frac{1}{\eta}=1.5$, $\frac{1}{\eta}=2.5$ and $\frac{1}{\eta}=3.5$. Thus, this shows that by decreasing $\eta$ the ground state energy increases.

\section{Scalar field in the cosmic string rainbow gravity}

In this section, we deal with the scalar field in the cosmic string spacetime in the context of the rainbow gravity. In the scenario of general relativity, the Klein-Gordon equation is given by \cite{Vitoria:2018its, Bezerra:1997mn}
\begin{eqnarray}
m^{2}\Phi=\frac{1}{\sqrt{-g}}\,\partial_{\mu}\left[\sqrt{-g}\,g^{\mu\nu}\,\partial_{\nu}\right]\Phi,
\label{2.1}
\end{eqnarray}
where $g=\mathrm{det}\left(g_{\mu\nu}\right)$ and $m$ is the mass of the relativistic particle. Then, from Eq. (\ref{1.4a}), we have $\sqrt{-g}=\frac{\eta\,r}{\left(1-\epsilon\,x^{2}\right)^{3/2}}$. Hence, with respect to the line element (\ref{1.4a}), the Klein-Gordon equation in the cosmic string spacetime in the context of the rainbow gravity is 
\begin{eqnarray}
m^{2}\Phi=-\frac{\partial^{2}\Phi}{\partial t}+\left(1-\epsilon\,y^{2}\right)\left[\frac{\partial^{2}}{\partial r^{2}}+\frac{1}{r}\frac{\partial}{\partial r}+\frac{1}{\eta^{2}r^{2}}\,\frac{\partial^{2}}{\partial\varphi^{2}}+\frac{\partial^{2}}{\partial z^{2}}\right]\Phi.
\label{2.2}
\end{eqnarray}

Since we have the cylindrical symmetry in this system, the solution to Eq. (\ref{2.2}) can be written as:
\begin{eqnarray}
\Phi\left(t,\,r,\,\varphi,\,z\right)=e^{-iEt+il\varphi+ikz}\,f\left(r\right),
\label{2.3}
\end{eqnarray}
where $l=0,\pm1,\pm2,\ldots$ and $k$ is a constant. In this way, by substituting Eq. (\ref{2.3}) into the Klein-Gordon equation (\ref{2.2}), we obtain the following equation for the function $f\left(r\right)$:
\begin{eqnarray}
f''+\frac{1}{r}\,f'-\frac{l^{2}}{\eta^{2}\,r^{2}}\,f+\lambda^{2}\,f=0,
\label{2.4}
\end{eqnarray}
where the parameter $\lambda$ has been defined in Eq. (\ref{1.16}), and we have also taken $k=0$. Note that Eq. (\ref{2.4}) is also the Bessel differential equation \cite{Arfken,Abramowitz}. Therefore, the solution to Eq. (\ref{2.4}) has the same form of Eq. (\ref{1.17}), i.e., 
\begin{eqnarray}
f\left(r\right)=A\,J_{\frac{\left|l\right|}{\eta}}\left(\lambda\,r\right)+B\,N_{\frac{\left|l\right|}{\eta}}\left(\lambda\,r\right),
\label{2.5}
\end{eqnarray} 
where $A$ and $B$ are also constants. The functions $J_{\frac{\left|l\right|}{\eta}}\left(\lambda\,r\right)$ and $N_{\frac{\left|l\right|}{\eta}}\left(\lambda\,r\right)$ are also the Bessel functions of first and second kinds \cite{Arfken,Abramowitz}, respectively. Since we search for a regular solution at the origin, thus, we need to take $B=0$ in Eq. (\ref{2.5}). Thereby, the radial wave function becomes
\begin{eqnarray}
f\left(r\right)=A\,J_{\frac{\left|l\right|}{\eta}}\left(\lambda\,r\right).
\label{2.6}
\end{eqnarray}

Next, we assume that the scalar particle is confined to a hard-wall confining potential. For a fixed radius $r_{0}$, then, we have the boundary condition:
\begin{eqnarray}
f\left(r_{0}\right)=0.
\label{2.7}
\end{eqnarray}
Let us also consider $\lambda\,r_{0}\gg1$, then, when $\frac{\left|l\right|}{\eta}$ is fixed, the Bessel function can be written as given in Eq. (\ref{1.20}). Thereby, by following the steps from Eq. (\ref{1.19}) to Eq. (\ref{1.21}), we obtain
\begin{eqnarray}
E_{n,\,l}=\pm\sqrt{\frac{m^{2}+\frac{\pi^{2}}{r_{0}^{2}}\left[n+\frac{\left|l\right|}{2\eta}+\frac{3}{4}\right]^{2}}{\left[1+\frac{\epsilon\,\pi^{2}}{E_{p}^{2}\,r_{0}^{2}}\left(n+\frac{\left|l\right|}{2\eta}+\frac{3}{4}\right)^{2}\right]}}. 
\label{2.8}
\end{eqnarray}

Hence, we have obtained in Eq. (\ref{2.8}) the spectrum of energy for a scalar field subject to a hard-wall confining potential in the modified background of the cosmic string spacetime. In contrast to Ref. \cite{Vitoria:2018its}, we can observe that the effects of rainbow gravity modify the relativistic energy levels for a scalar particle subject to a hard-wall confining potential. The effects of the rainbow gravity yield the term $\left[1+\frac{\epsilon\,\pi^{2}}{E_{p}^{2}\,r_{0}^{2}}\left(n+\frac{\left|l\right|}{2\eta}+\frac{3}{4}\right)^{2}\right]^{-1}$ inside the square root in Eq. (\ref{2.8}). We can also see that the scalar field does not interact with the topological defect, but the relativistic energy levels (\ref{2.8}) depend on the parameter $\eta$ of cosmic string spacetime. In this case, the analogue of the Aharonov-Bohm effect for bound states is yielded by the presence of the effective angular momentum $l_{\mathrm{eff}}=l/\eta$.

Note that with $\eta\rightarrow1$, then, the relativistic energy levels (\ref{2.8}) becomes
\begin{eqnarray}
E_{n,\,l}=\pm\sqrt{\frac{m^{2}+\frac{\pi^{2}}{r_{0}^{2}}\left[n+\frac{\left|l\right|}{2}+\frac{3}{4}\right]^{2}}{\left[1+\frac{\epsilon\,\pi^{2}}{E_{p}^{2}\,r_{0}^{2}}\left(n+\frac{\left|l\right|}{2}+\frac{3}{4}\right)^{2}\right]}}. 
\label{2.9}
\end{eqnarray}

Therefore, we have in Eq. (\ref{2.9}) the relativistic energy levels for the scalar field subject to a hard-wall confining potential in the modified background of the Minkowski spacetime. We can see the influence of the rainbow gravity on the relativistic energy levels (\ref{2.9}) through the term $\left[1+\frac{\epsilon\,\pi^{2}}{E_{p}^{2}\,r_{0}^{2}}\left(n+\frac{\left|l\right|}{2}+\frac{3}{4}\right)^{2}\right]^{-1}$ inside the square root. Note that by taking $\epsilon\rightarrow0$ in Eqs. (\ref{2.8}) and (\ref{2.9}), we obtain the relativistic energy levels analogous to that obtained in Ref. \cite{Vitoria:2018its} in the absence of rainbow gravity.

Let us now end this section by pointing out some important features of our results. First of all, the behaviour of the energy levels (\ref{2.8}), for $n=l=0$, is very similar to that one plotted in Figs. \ref{f1} and \ref{f2}. The only difference between the energy levels in Eq. (\ref{1.21}) and that one in (\ref{2.8}) is the spin present in the parameter $\nu$ defined in Eq. (\ref{1.16}).

We should also mention that for high values of the quantum numbers $l$ and $n$ the energy levels (\ref{1.21}) and (\ref{2.8}) tends to the finite value $E\simeq \epsilon^{-\frac{1}{2}}E_p$. That is, in the rainbow gravity approach, the Planck energy sets a natural cut off for the energy, in contrast to the normal situation without rainbow gravity. In this case, as it is known, the energy diverges. Therefore, one of the several features present in the rainbow gravity approach is the elimination of possible divergences, as in our case.

\section{Conclusions}

We have studied the Dirac and the scalar fields in the cosmic string spacetime in the context of rainbow gravity. We have chosen a scenario of the rainbow gravity determined by the rainbow functions: $g_{0}\left(y\right)=1$ and $g_{1}\left(y\right)=\sqrt{1-\epsilon\,y^{2}}$ (see Eq. (\ref{1.4})). Then, by analysing a particular case of the confinement of the Dirac particle to a hard-wall confining potential in a modified background of the cosmic string spacetime, we have seen that the spectrum of energy is modified due to the effects of the rainbow gravity in contrast to the spectrum of energy obtained in Ref. \cite{Bezerra:1997mn} in the cosmic string spacetime. Besides, we have seen an analogous effect to the Aharonov-Bohm effect for bound states \cite{ABeffect,Bezerra:1997mn} due to the effects of the topology of the cosmic string. We have also analysed the limit case where $\eta\rightarrow1$. In this case, we have obtained in Eq. (\ref{1.22}), the relativistic energy levels for the Dirac field subject to a hard-wall confining potential in the modified background of the Minkowski spacetime.

Furthermore, by analysing the scalar field subject to a hard-wall confining potential in the same scenario of rainbow gravity, i.e., the scenario determined by the rainbow functions established in Eq. (\ref{1.4}), we have observed that the relativistic spectrum of energy is modified by comparing with Ref. \cite{Vitoria:2018its}. This change in the spectrum of energy stems from the effects of the rainbow gravity. We have also seen a relativistic analogue of the Aharonov-Bohm effect for bound states, since there is the influence of the topology of the cosmic string on the relativistic spectrum of energy. Finally, we have taken $\eta\rightarrow1$, and thus, we have obtained in Eq. (\ref{2.9}) the relativistic spectrum of energy for the scalar field subject to a hard-wall confining potential in the modified background of the Minkowski spacetime.

We have also plotted in Figs. \ref{f1} and \ref{f2} the dimensionless ground state energy and shown the clear difference between the case with and without the rainbow gravity. The behaviour presented in Figs. \ref{f1} and \ref{f2} is essentially the same for both the fermionic and scalar cases. In addition, we have pointed out that the rainbow gravity introduces a natural cut off for the energy when the quantum numbers $\ell$ and $n$ take high values.

It is worth bringing attention to the quantum effects that can rise from a topological defect spacetime with torsion in the framework of rainbow gravity. Recently, one of us has discussed quantum effects in the spacetime with a screw dislocation \cite{Vitoria:2018its,sb,sb2} and in the spacetime with a spiral dislocation \cite{mb}. Therefore, this work raises the possibility of investigating the effects of the rainbow gravity in several scenarios related with the interface between general relativity and relativistic quantum mechanics \cite{put,int1,int2,int3,int4,int5,int6,int7,int8,int9,int10,int11,int12,int13,int14,extra3,extra4,extra5,extra6,1,5,8,9,10,11,12,14,15,16,17,18,19,20,21,22,23,24,25,27,28,29,30,31}.

\acknowledgments{The authors would like to thank CNPq for financial support.}

\end{document}